\magnification=1200
\baselineskip=18truept
\input epsf

\def\preprint{Y}

\def\cap{\hsize=4.6in}
\def\figure#1#2#3{\if \preprint Y \midinsert \epsfxsize=#3truein
\centerline{\epsffile{figure_#1_eps}} \halign{##\hfill\quad
&\vtop{\parindent=0pt \hsize=4.6in \strut## \strut}\cr {\bf Figure
#1}&#2 \cr} \endinsert \fi}

\def\captionfiveone{\cap
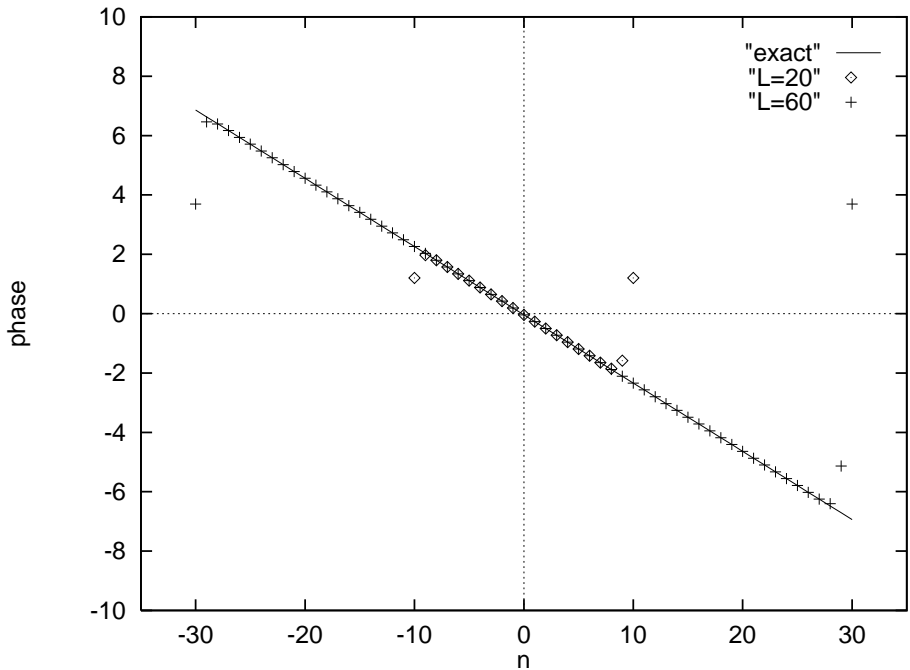
$D^u_{\rm lat}(0.23,0.37+n_2;L)$ as a function of $n_2$ for
$L=20$ and $L=60$ along with the continuum 
result from section 3 (cf eqn(3.1)).}
\def\captionfivetwo{\cap Locus of $D^u_{\rm lat}(0.37,h_2;L)$ as
$h_2$ is varied from $0$ to $0.5$ for $L=10,18$ along with the continuum
result from section 3(cf eqn(3.1)).}
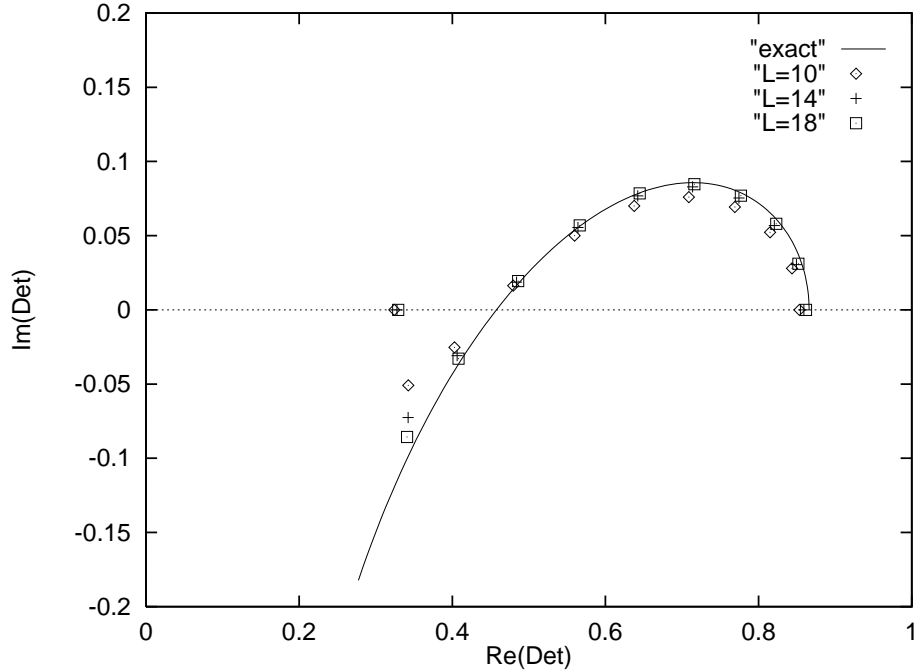
\def\captionfivethree{\cap Locus of $D^s_{\rm lat}(0.37,h_2;L)$ as
$h_2$ is varied from $0$ to $0.5$ for $L=10,14,18$ along with the continuum
result from section 3(cf eqn(3.1)).}
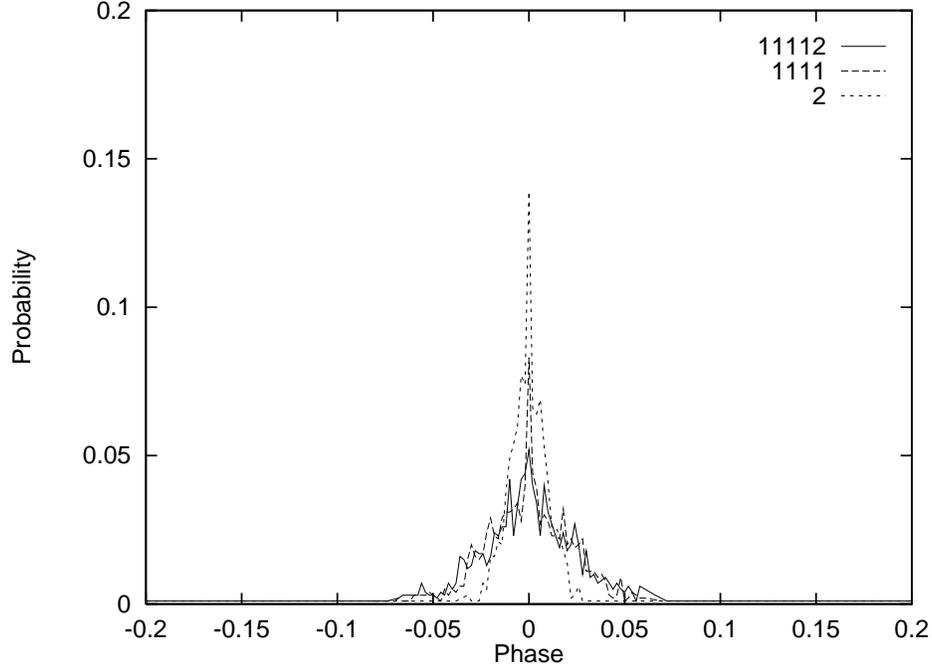
\def\captionsixone{\cap Normalized phase distribution 
for a sample gauge field
at $eL=0.1\pi$ with $h_1=h_2=0$ on a $6\times 6$ lattice.}
\def\captionsixtwo{\cap Normalized phase distribution 
for a sample gauge field
at $eL=0.1\pi$ with $h_1=h_2=0$ on a $8\times 8$ lattice.}
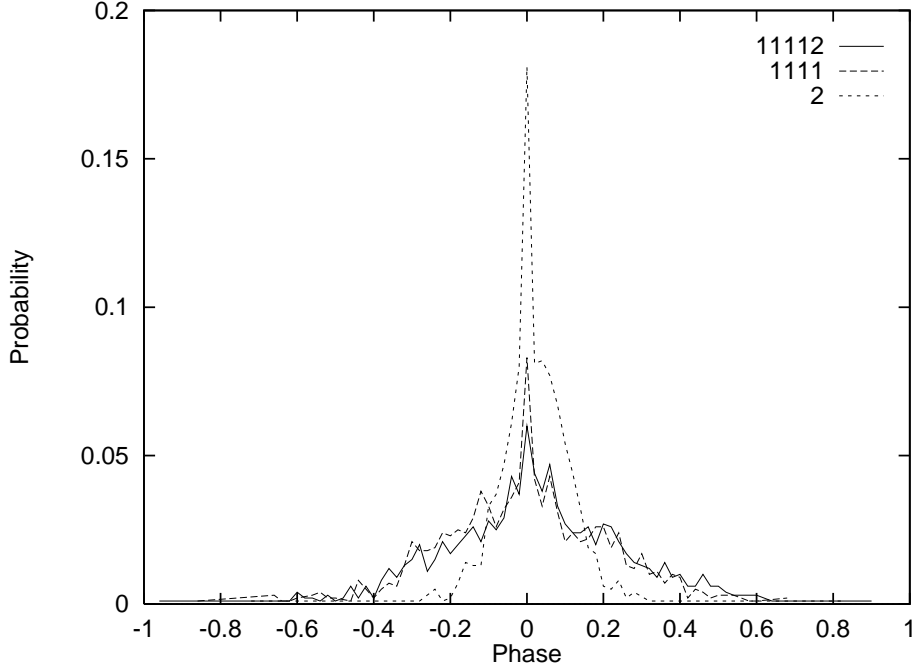
\def\captionsixthree{\cap Normalized phase distribution 
for a sample gauge field
at $eL=\pi$ with $h_1=h_2=0$ on a $6\times 6$ lattice.}
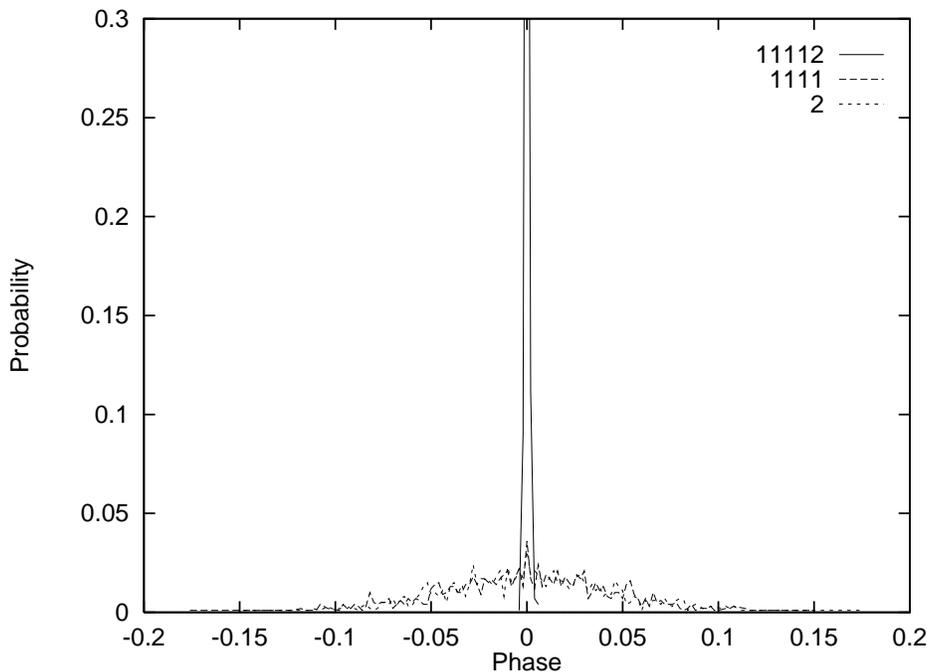
\def\captionsixfour{\cap Normalized phase distribution 
for a sample gauge field
at $eL=\pi$ and $\kappa=0.1$ with $h_1=h_2=0$ on a $6\times 6$ lattice.}
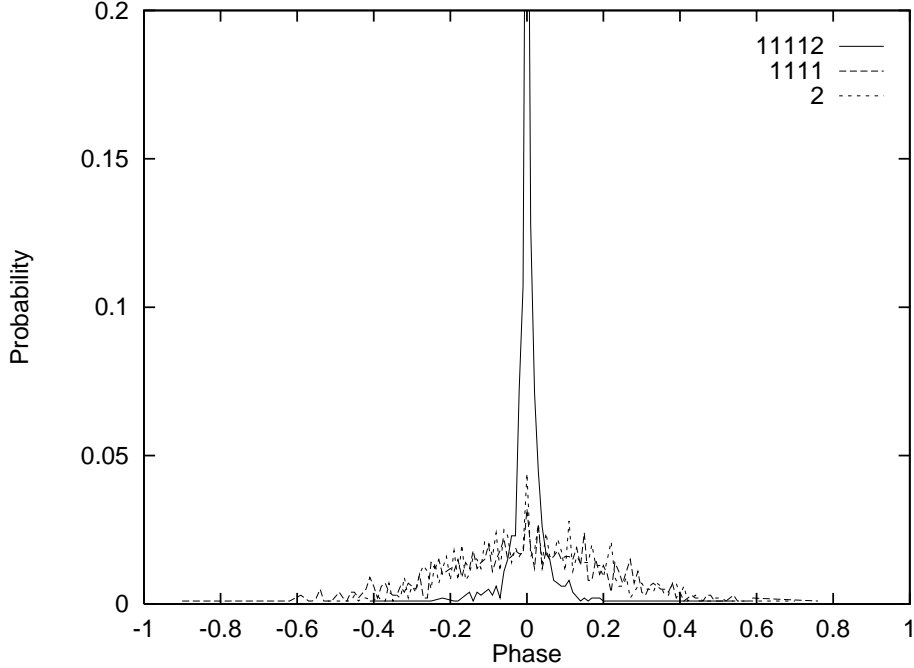
\def\captionsixfive{\cap Normalized phase distribution 
for a sample gauge field
at $eL=\pi$ and $\kappa=0.5$ with $h_1=h_2=0$ on a $6\times 6$ lattice.}
\def\captionsixsix{\cap Normalized phase distribution 
for a gauge field with
no electric field ($\phi=0$) and $h_1=0.07$, $h_2=0.13$ on a 
$6\times 6$ lattice
.}
\def\captionsixseven{\cap Normalized phase distribution 
for a gauge field with
no electric field ($\phi=0$) and $h_1=0.23$, $h_2=0.37$ 
on a $6\times 6$ lattice
.}
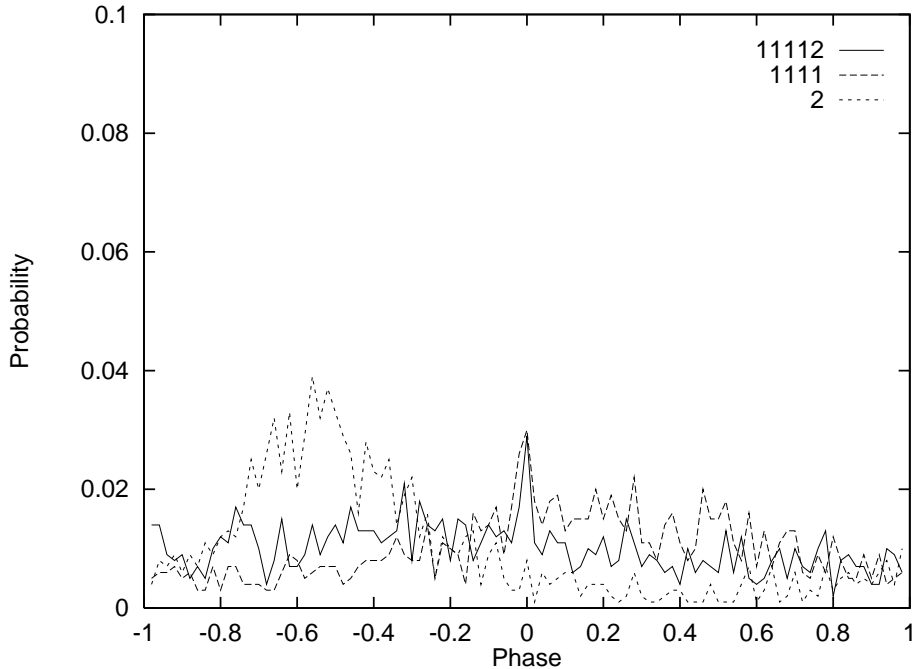
\def\captionsixeight{\cap Normalized phase distribution 
for a sample gauge field
at $eL=\pi$ with $h_1=0.23$, $h_2=0.37$ on a $6\times 6$ lattice.}
\def\captionsixnine{\cap Normalized phase distribution for a 
gauge field with
no electric field ($\phi=0$) and $h_1=0.23$, $h_2=0.37$ 
when the points on the gauge orbit are restricted to be constant on blocks 
($L=9$, $L_b=3$) and
the starting point is the ``uniform'' configuration.}
\def\captionsixten{\cap Normalized phase distribution 
for a gauge field with
no electric field ($\phi=0$) and $h_1=0.23$, $h_2=0.37$ 
when the points on the gauge orbit are restricted to be constant on blocks 
($L=9$, $L_b=3$) and
the starting point is the ``singular'' configuration.}
\def\captionsevenone{\cap Normalized phase distribution 
for a gauge field with
no electric field ($\phi=0$) and $h_1=0.07$, $h_2=0.13$ 
on a $6\times 6$ lattice.
}
\def\captionseventwo{\cap Normalized phase distribution 
for a gauge field with
no electric field ($\phi=0$) and $h_1=0.23$, $h_2=0.37$ 
on a $6\times 6$ lattice.
}

\def\chiral{{\rm{\bf C}}}
\def\wilson#1{{\rm{\bf B^#1}}}
\def\ham#1{{\rm{\bf H^#1}}}

\def\inter{1}
\def\sw{2}
\def\nna{3}
\def\fnn{4}
\def\detu{5}
\def\anom{6}
\def\nnb{7}
\def\patch{8}
\def\nnv{9}
\def\shsm{10}

\line{\hfill UW/PT--96--04}
\line{\hfill DOE/ER/40561-254-INT96-00-123}
\line{\hfill RU--96--18}
\vskip 2truecm
\centerline{\bf Anomaly free U(1) chiral gauge
theories on a two dimensional torus}

\vskip 1truecm
\centerline{Rajamani Narayanan}
\centerline{\it Institute for Nuclear Theory, Box 351550}
\centerline{\it University of Washington, Seattle, WA 98195-1550.}  
\vskip .1in
\centerline{Herbert Neuberger}
\centerline{\it Department of Physics and Astronomy}
\centerline{\it Rutgers University, Piscataway, NJ 08855-0849. }
\vfill
\centerline{\bf Abstract}
\vskip 0.75truecm
We consider anomaly free
combinations of
chiral fermions coupled to $U(1)$ gauge fields on a 2D torus
first in the continuum and then on the lattice in the overlap formulation.
Both in the continuum and on the lattice, when the background consists
of sufficiently large
constant gauge potentials the action induced by the fermions
varies significantly
under certain singular gauge transformations.
``Ruling away'' such discontinuities cannot be justified in the
continuum framework and does not naturally fit on the lattice.
Complete gauge invariance in the continuum can be restored
in some models by choosing special boundary 
conditions for the fermions.
Evidence is presented that gauge averaging the overlap phases in these
models produces correct continuum results.
\vfill
\eject
\centerline{\bf 1. Introduction}
\medskip

Two dimensional chiral gauge theories provide
a convenient testing ground for non-perturbative regularizations since
they are much simpler to simulate numerically than four dimensional ones.
As in four dimensions, there are anomalies in two dimensions that
have to be cancelled and anomaly free theories can have global charges that
are broken by topologically non-trivial gauge fields.
This paper is about abelian chiral gauge theories with several
Weyl fermions in Euclidean space--time.
The left handed Weyl fermions, labeled by $j$,
have charge $q_{Lj}$
and the right handed ones, labeled by $i$, have charge $q_{Ri}$.
Anomalies are cancelled by requiring $\sum_i q^2_{Ri} = \sum_j q^2_{Lj}$.
In order to avoid infra-red problems
we consider these theories on a
compact manifold; we prefer the gravitationally flat background
of the torus. Invariance under rotations
by 90 degrees is maintained by setting both sides of
the torus equal to a common length, $l$.

Asymptotic freedom implies that perturbation theory works well
when $l$ is small relative to the model's intrinsic scales. On small 
tori the field strength is also 
small, so the gauge potentials are
essentially constant up to gauge transformations.
Further, the probability for nonzero topology is 
suppressed.

In the classical limit the gauge fields
are connections on a bundle over the torus.
For our purposes it is convenient to
describe the torus utilizing four
square patches of identical size.
Each patch intersects every other patch in two separate 
intersections.
The gauge fields defined locally on the different patches
are gauge related on their intersections by transition
functions.

The quantum gauge fields are close to a classical connection
on a trivial bundle with zero curvature. The gauge invariant
content of this configuration is expressed by two phases (angles),
$e^{2\pi i h_\mu }, \mu =1,2$. 
One may associate these phases with two
Polyakov loops wrapping once around the torus in the two directions. 
We consider two gauges: 
In the
``uniform'' gauge the transition functions are all unity and all 
connections are constant. 
We have
$A_\mu = 2\pi h_\mu / l, \mu=1,2$ uniformly on all patches. In
the ``singular'' gauge the connections are all zero and the
phases reside entirely in the
transition functions. The transition functions are nontrivial on two
narrow strips winding around the torus in two directions 
and are chosen to produce the correct phase factor for
parallel transport along any loop.

Consider one of the several Weyl fermions in any of the above
backgrounds. The associated Grassmann path integration 
generates an effective action that depends 
on the gauge field. The perturbative anomaly vanishes since the 
curvature is zero. On a torus
the set of continuous $U(1)$
gauge transformations splits into an infinite
collection
of disconnected pieces labeled by the number of
windings in the two directions. The absence of
a perturbative anomaly
does not ensure invariance under winding gauge transformations.
Consider the winding gauge transformations which map a ``uniform''
gauge $A_\mu = 2\pi h_\mu /l$ to another ``uniform'' gauge
$A_\mu$ with the $h_\mu$ shifted by some integers.
Such a gauge transformation
does not affect the ``singular'' gauge representation.
Therefore, if the effective action
is defined in the ``singular'' gauge, gauge invariance under
these winding gauge transformations is automatic. However, if the
effective action is defined in the ``uniform'' gauge, gauge
invariance under the winding gauge transformations is not guaranteed.
It turns out that if the Feynman diagrams are
summed in the ``uniform'' gauge (using zeta-function regularization
for example), the effective action
indeed changes under a winding gauge transformation. (Any
effective action that is smooth in the constant gauge potentials,
that obeys some natural discrete symmetries and that reproduces 
all convergent
Feynman diagrams will end up violating the winding symmetries.)
When the contributions from all the Weyl fermions are
summed up invariance under the winding gauge transformations is restored
as a consequence of $\sum_i q^2_{Ri} = \sum_j q^2_{Lj}$. 
The winding gauge transformations do
not directly violate gauge invariance
in either the ``singular'' or the ``uniform'' gauges.
However, the imaginary part of the 
the effective action in the ``singular'' gauge differs from the one
in the ``uniform'' gauge. The
gauge transformation connecting a ``singular'' background
to a ``uniform'' one is singular.

We concluded 
that even after perturbative
anomalies have been cancelled some gauge non--invariance still
afflicts the effective action.
We used several patches to describe the gauge fields
although there is no non-trivial topology. As physicists, we
wish to rephrase the description employing no patches.
Insisting on using functions defined over the whole torus,
we end up describing
the ``singular'' gauge configurations by potentials
$A_\mu (x)$ that have linear $\delta$-function singularities in
the $\mu$ direction. As a result, the dimension of the gauge
field part of the Dirac operator equals that of the derivative
part and new effects can arise when the ultraviolet
behavior is regulated.
In a purely vectorial theory the problem disappears
since one can preserve exact 
pairwise phase cancelation between left and right movers.
The problem disappears even in the chiral case if at least
one of the phase factors is exactly unity, or if both
phase factors are in some limited region around unity (the
actual size of the region is model dependent).
In summary, the ingredients necessary to
observe the gauge invariance violation
are:
\item{$\bullet$} A gauge background that has
two nontrivial $e^{2\pi i h_\mu }$'s
with values not too close to unity.
\item{$\bullet$} A gauge transformation that has a
singularity whose effect is to concentrate the parallel
transporters around the torus to a line transverse to the loop.
There is nothing ``random'' about these
gauge transformations. They are very ``ordered'' and perfectly
reasonable even in a continuum setting.

In the continuum one may try to forbid ``singular'' gauge fields.
This would be somewhat artificial,
since it is not enforced
by the action and since it would be unnecessary in the vector case.
Such a decree makes little sense quantum mechanically, where the
domain of integration in the functional integral
is not restricted by ordinary
smoothness properties.
On a lattice, one lets the action govern the type of gauge configurations
that appear and there is no natural way of imposing a restriction of this sort.
The above problem will be encountered in any lattice regularization
that avoids non-local gauge fixing.
For example,
if we used interpolating methods, following [\inter], gauge singularities 
of the linear $\delta$-function type are abundant. If they are
to be allowed the singularities we are focusing on certainly cannot
be ruled out.
Alternatively one may wish to restrict the phase factors,
$e^{2\pi i h_\mu}$, 
to a range
sufficiently close to unity.
This does not seem consistent with translational invariance:
In the computation of condensates in vectorial theories in topologically
nontrivial backgrounds [\sw] the quantities $lh_\mu$ play the role
of center of mass collective coordinates
and one needs to integrate over the entire range of each $h_\mu$
in order to restore translational invariance. So, as long as we are
on a torus, there seems to be no natural escape. However, on a sphere,
or cylinder, it is possible that the problem can be avoided.

The above problem is a generic feature in the continuum
and any reasonable regulator should reproduce the
violation of gauge invariance. We shall show in some detail that
the overlap formulation [\nna] has gauge violations
in precisely the circumstances dictated by the continuum.
A considerable effort has gone into convincing
ourselves that these are, essentially, the {\it single} situations
where consequential
gauge breaking occur in the overlap. In particular, just
the mere randomness of the gauge fluctuations
creates no difficulties.
On the contrary, according to
F{\" o}rster, Nielsen and Ninomiya [\fnn]
random gauge fluctuations will restore 
gauge invariance at long distances
in any model with relatively mild
gauge breaking in the ultraviolet.

Originally we planned
to test the overlap non--perturbatively
for abelian chiral models defined on the torus [\nna].
The
present paper shows that most models do
not provide a clean testing ground.
Nevertheless, our results confirm the
overlap as a valid regularization of chiral gauge theories.
In particular, we construct a special class of
abelian chiral models on the torus where gauge breaking
are completely eliminated in the continuum. This is
achieved by a judicious choice of fermionic boundary
conditions. The lattice overlap reproduces the good
continuum behavior correctly.
If we replaced the boundary conditions in these
models by simpler prescriptions 
a non--discriminating lattice test would indicate some persistent
gauge breaking effects which could be misinterpreted as a lattice
problem. We wish to stress again our view that these effects 
are not a lattice deficiency but a continuum one.

The paper is organized as follows. In section 2,
a specific chiral $U(1)$
model is defined in the continuum with anti-periodic boundary
conditions for all fermions. In section 3,
the chiral determinant is studied in the presence of constant
gauge fields and the associated gauge transformations. Explicit examples
of gauge field configurations show that the theory is not
gauge invariant unless the zero modes of the gauge fields are sufficiently
small.
In section 4, the overlap
formulation on the lattice
of the theory in section 2 is described.
In section 5, we show that the overlap reproduces the 
lack of gauge invariance
found in section 3.
In section 6 we discuss extensively gauge averaging along orbits.
We assess the statistical significance
of the violation of gauge invariance and investigate whether
relatively large phase factors in
conjunction with singular gauge transformations are its main source. 
Our objective is to show that employing gauge averaging along
orbits as a way to restore exact gauge invariance on the lattice
is a valid method. As long as there are no gauge violations in
the continuum one recovers the desired outcome, and, reassuringly,
if such gauge violations do exist, a clear signal is obtained.
In section 7 we change the chiral model by
picking new boundary conditions
for the fermions. Now complete
gauge invariance holds in the continuum and the overlap formalism 
in conjunction with gauge averaging
produces correct results on all gauge orbits.
In section 8 we
summarize our findings and draw some conclusions.
Readers who prefer a slightly more detailed 
overview 
now should skip to section 8 before 
reading sections 2 through 7.

\bigskip
\centerline{\bf 2. A 11112 model}
\medskip
For definiteness we consider a specific $U(1)$ chiral model. 
All results
can be carried over to any other $U(1)$ chiral model.
We pick $q_{Li}=1, i=1,2,3,4$
and a single right handed fermion with charge $q_{R}=2$.
Numerically, it is more economical to deal with this model than
with the more traditional ``345'' model. The global chiral
$SU(4)$ symmetry among the left-handed fermions in the ``11112''
model is of potential
interest while the 345 model has
no global nonabelian symmetry.
Also, the fermion number violating 't Hooft vertex is slightly
simpler in our case than in the 345 model [\nna].

The partition function of our 11112 model is given by
$${\cal Z} = \int [dA_\mu][d\bar\psi][d\psi] \exp (-S[\bar\psi,\psi,A_\mu])
\eqno{(2.1)}$$
$$S[\bar\psi,\psi,A_\mu] = S_g[A_\mu] + S_f[\bar\psi,\psi,A_\mu]\eqno{(2.2)}$$
$$S_g[A_\mu] = {1\over 4e^2} \int d^2 x F^2_{\mu\nu}\eqno{(2.3)}$$
$$S_f[\bar\psi,\psi,A_\mu]=
-\sum_{k=1}^4\int d^2 x \bar\psi_k^L \sigma_\mu (\partial_\mu+iA_\mu )\psi_k^L
-\int d^2 x \bar\psi^R \sigma^*_\mu (\partial_\mu+2iA_\mu )\psi^R\eqno{(2.4)}
$$
$\sigma_1=1$ and $\sigma_2=i$ in (2.4). The base manifold is an
$l\times l$ torus and the fermion
fields obey anti-periodic boundary conditions.

$e$ and $A_\mu$ have dimensions
of mass. The field $F_{\mu\nu} = \partial_\mu A_\nu - \partial_\nu A_\mu
\equiv \epsilon_{\mu\nu} E$ has dimensions of mass
squared and the topological
invariant $Q_T \equiv {1\over 2\pi} \int d^2 x E $ can take any integer value.
The quantization of $Q_T$ gives a precise meaning to the
magnitudes of the charges assigned to the fermions.
The path integral includes a sum over all topological sectors
but, as outlined in the introduction,
we will restrict ourselves to $Q_T =0$.

In the zero topological sector we
use functions to describe the gauge potentials.
The gauge fields are decomposed as
$$A_\mu = \epsilon_{\mu\nu}\partial_\nu \phi  - i g^*(x) \partial_\mu g(x)
+ {2\pi\over l} h_\mu\eqno{(2.5)}$$
$h_1$ and
$h_2$ are zero modes of the gauge fields, i.e.,
$h_\mu = {1\over 2\pi l} \int d^2x A_\mu$.  The electric field density is
$$E(x)=\partial^2 \phi(x).\eqno{(2.6)}$$
Given $E(x)$, (2.6) can be used to solve for $\phi(x)$.  $\phi(x)$
is a periodic function on the torus with a vanishing zero mode.
$g(x)$ is a $U(1)$ valued function.
$h_1$, $h_2$ and $E(x)$ constrained by $Q_T =0$ 
constitute the gauge invariant
content of the gauge field. 
The integral over $A_\mu$ can be replaced by an
integral over $\phi(x)$, $h_1$, $h_2$ and $g(x)$.
The first three label different
gauge orbits and the last one labels points on a gauge orbit.

Only small and relatively slowly varying
$\phi$ fields are allowed by $S_g$. However, 
$S_g$ is independent of the
$h_\mu$'s and has no control over their
fluctuations. The probability distribution of the $h_\mu$'s
is only influenced by the fermions through the induced effective action.
In a finite Euclidean volume there is no mechanism to ``freeze'' these
variables.\footnote{*}{ For example, if one direction were infinite
one of the $h_\mu$'s would have to minimize the effective potential
and the problematic gauge backgrounds would disappear.}
Therefore, once we establish that for some range of $h_\mu$'s
we have a problem with gauge invariance, the issue cannot
be discounted on the grounds that there is no weight for these
backgrounds.

$g(x)$ itself is an arbitrary function, no condition on its magnitude
or variability is imposed; in other words we include gauge 
singularities. As described in the introduction
one can always split
the torus into patches so that $g(x)$ is smooth in each patch. Discontinuities
in $g(x)$
between patches can be stored in the transition functions between the
patches. The Dirac-Weyl operators act on sections and are 
smooth mappings
of sections into other sections on the same coordinate bundle. 
We also can define a perfectly reasonable eigenvalue 
problem
for the Weyl-Dirac operator
although the eigenvalues are not scalars under rotations. In summary,
allowing $g(x)$ to have discontinuities
is both reasonable and mathematically
acceptable. Of course, defining the chiral {\it determinants}
requires care in general,
and even more so when gauge singularities are present. But this is the
task of renormalized quantum field theory.

\bigskip
\centerline{\bf 3. Chiral determinant when $\phi=0$}
\medskip
For $\phi=0$, $S_g=0$ and all values of $h_\mu$ along with all their
possible gauge transforms are equally likely for the pure gauge action.
In this section we study the chiral determinant when $\phi=0$. We start with
$A^u_\mu(x)={2\pi i \over l}h_\mu$.
The superscript $u$ means that $A_\mu$ is in the ``uniform'' gauge.
The $h_\mu$'s in the range $[-1/2,1/2)$ are gauge inequivalent.
All other values of $h_\mu$ can be obtained by gauge transformations.
Apart from the gauge transformations that take
$h_\mu \rightarrow h_\mu+n_\mu$, we will also consider gauge transformations
that result in the singular gauge field configuration,
$A^s_\mu(x)=2\pi i h_\mu \delta(x_\mu )$ (no sum on $\mu$).
Let $D^u(h_1,h_2)$ and $D^s(h_1,h_2)$ denote the chiral determinant
of a single left handed Weyl fermion in the two backgrounds $A^u_\mu$
and $A^s_\mu$.

The result for $D^u$ is already known [\detu].
It can be obtained by zeta function
regularization and other continuum methods.\footnote{**}
{In zeta-function regularization it is easiest to obtain
an answer for the $h_\mu$ restricted to $[-1/2,1/2)$ and then
the answer is holomorphic in $h_1+ih_2$. To restrict
the gauge violation to the imaginary part of the effective
action one needs to add a real quadratic term which 
breaks holomorphy.}
Although the electric field vanishes, $D^u$ is anomalous, not being 
invariant under $h_\mu\rightarrow h_\mu+n_\mu$. 
The explicit formula for
$D^u$ is
$$\eqalign{&{D^u(h_1,h_2)\over D^u(0,0)}=\cr & e^{i\pi h_2 (h_1 +ih_2 )}
{{\prod_{n=1}^{\infty} [(1+e^{-2\pi (n-1/2) -2\pi i h_1 + 2\pi h_2 })
(1+e^{-2\pi (n-1/2) + 2\pi i h_1 - 2\pi h_2 })]}\over
{\prod_{n=1}^{\infty} [(1+e^{-2\pi (n-1/2)})^2]}} \cr}
\eqno{(3.1)}$$
{}From (3.1) one obtains 
$$D^u(h_1+n_1,h_2 +n_2 )= 
e^{i\pi (n_1 h_2 -n_2 h_1 )} D^u(h_1,h_2)\eqno{(3.2)}$$
The lack of gauge invariance arises because the determinant 
was made 
a smooth function of $h_1$ and $h_2$. Requiring smoothness 
for the uniform background is natural 
since the fermions see locally
a field that depends linearly on $h_1$ and $h_2$. 
Also, in a finite Euclidean volume there 
is no thermodynamic mechanism to induce a singularity.

Now let us turn our attention to $A^s(x)$ and the associated $D^s$.
Again one can compute the determinant by Feynman diagram methods.
For $h_1$ and $h_2$ close to zero we expect the result to match 
$D^u$, but it should not change under 
$h_\mu\rightarrow h_\mu+n_\mu$ because the fermion sees only
the transition functions $e^{2\pi i h_\mu }$. 
We conclude that in the fundamental domain
$$D^s (h_1,h_2) = D^u (h_1,h_2) \ \ \ \ \ \forall h_\mu \ \ \ \
-{1\over 2} \le h_1, h_2 < {1\over 2} \eqno{(3.3)}$$
and, for all integers $n_\mu$,
$$D^s(h_1+n_1,h_2+n_2) = D^s (h_1,h_2).\eqno{(3.4)}$$

The formulae for $D^u$ and $D^s$ do not
yield a gauge invariant determinant for the 11112 model.
In a uniform background field the partition function 
of all the fermions is
$$Z^u(h_1,h_2) = \Bigl [ D^u(h_1,h_2) \Bigr ]^4
                \Bigl [D^u(2h_1,2h_2) \Bigr ]^*\eqno{(3.5)}$$
{}From (3.2) is follows that
$$Z^u(h_1+n_1,h_2+n_2) = Z^u(h_1,h_2) \eqno{(3.6)}$$
showing that the effective action restricted to $A^u$ is gauge invariant.
The disappearance of the violation in (3.2) is due to the
anomaly cancelation condition $1^2 +1^2 +1^2 +1^2 = 2^2 \equiv 4$.
For the singular gauge field the partition function is 
$$Z^s(h_1,h_2) = \Bigl [ D^s(h_1,h_2) \Bigr ]^4
                \Bigl [D^s(2h_1,2h_2) \Bigr ]^*  \eqno{(3.7)}$$
{}From (3.4) it follows that (3.7) also is gauge invariant.
Let us compare
the partition function for the singular field 
with that for the uniform field.
Using (3.3) we replace (3.7) by 
$$Z^s(h_1,h_2) = \Bigl [ D^u(f(h_1),f(h_2)) \Bigr ]^4
                      \Bigl [D^u(f(2h_1),f(2h_2)) \Bigr ]^* ,
                \eqno{(3.8)}$$
where $f(h)$ is the function that takes $h$ into the range $[-1/2,1/2)$
by an appropriate shift by an integer.
Only if $f(2h_\mu) = 2f(h_\mu)$ for $\mu=1,2$ do we have
$$Z^s(h_1,h_2)=Z^u(h_1,h_2)\eqno{(3.9)}$$
(3.9) will hold if the $h_\mu$ are each in the segment $[-1/4,1/4]$.
The result is gauge invariant if the $h_\mu$
are sufficiently small but not if they are large.

The conclusion we arrived at is somewhat unexpected. It holds 
for any anomaly free chiral $U(1)$ model on the torus
as long as all fermions obey anti-periodic
conditions.
The region of $h_\mu$
where 
gauge invariance is preserved will be a function of the model
but there will always exist ranges of $h_\mu$ where gauge
invariance is lost. 
In the following sections we will show that this result holds
rigorously in the overlap regularization. Of course, many other
regularizations would also reproduce the same effect.

\bigskip
\centerline{\bf 4. Overlap formalism}
\medskip
We embed an $L\times L$ lattice in the continuum $l\times l$ torus. 
The
lattice spacing $a$ is given by $aL=l$. The parallel transporters
$$U_\mu (n)= e^{i\int_0^1 A_\mu (x+ta\hat\mu) dt}\eqno{(4.1)}$$
on the links of the lattice are constructed out of $\phi$, $g$ and
$h_\mu$ as follows. 
Replacing the continuum $g(x)$ we attach to each 
lattice site $n$ a $U(1)$ group variable $g(n)$. 
to the plaquette with corners at 
$n$, $n+\hat\mu$, $n+\hat\nu$ and $n+\hat\mu+\hat\nu$ we associate
an angle $\phi(n)$, a discretization of the continuum $\phi(x)$.
The $\phi(n)$'s are restricted by the condition $\sum_n \phi(n)=0$.
Then
$$U_\mu(n)= g(n+\hat\mu) g^*(n)
            e^{i\epsilon_{\mu\nu}[\phi(n)-\phi(n-\hat\nu)]}
            e^{i{2\pi\over L}h_\mu}\eqno{(4.2)}$$
It is best to visualize the $\phi$
variables as living on the sites of the dual lattice.
The Wilson gauge action is
$$ \eqalign{
S^w_g= & {1\over e^2}
       \sum_n {\rm Re} [ 1- U_2(n) U_1(n+\hat 2) U^*_2(n+\hat 1) U^*_1(n) ]
       \cr
=    &  {1\over e^2}
       \sum_n
       [1-\cos ((\partial_1^*\partial_1 + \partial_2^*\partial_2 ) \phi(n))]
       \cr } \eqno{(4.3)}$$
where $\partial_\mu$ are the forward lattice derivatives and $\partial^*_\mu$
are the backward lattice derivatives.
The electric field per plaquette is
$E(n) = (\partial_1^*\partial_1 + \partial_2^*\partial_2 ) \phi(n)$.
The gauge invariant variable is the plaquette parallel transporter,
$e^{iE(n)}$.
The $h_\mu$ can be restricted to $[-1/2, 1/2)$ 
since the $g(n)$'s are arbitrary.

The fermionic path integral is defined on the lattice using 
the overlap formalism [\nna].
The continuum fermionic determinant for a
left handed Weyl fermion
with $q_L=1$ is replaced by an inner product of two states:
$$\int d\bar\psi_L  d\psi_L e^{-S_f(\bar\psi_L,\psi_L,A_\mu)} =
{}^{\rm WB}_{\ U}\!\!<L-|L+>^{\rm WB}_U \eqno{(4.4)}$$
The states $|L\pm>^{\rm WB}_U$
are the ground states of two many body Hamiltonians
$$
{\cal H}^\pm =\sum_{n\alpha,m\beta} a^\dagger_{n,\alpha }
\ham\pm(n\alpha,m\beta; U)
a_{m,\beta },~~
\{a^\dagger_{n,\alpha }, a_{m,\beta }\} =\delta_{\alpha,\beta}\delta_{nm}
\eqno{(4.5)}$$
$\alpha,\beta=1,2$ and $n=(n_1 , n_2 )$ with
$n_\mu =0,1,2,.....,L-1 $. The single particle hermitian
hamiltonians $\ham\pm$ are given by:
$$\eqalign{
\ham\pm & =\pmatrix {\wilson\pm &\chiral\cr \chiral^\dagger&-\wilson\pm},\cr
\chiral(n,m) & ={1\over 2}
\sum_\mu \sigma_\mu (\delta_{m,n+\mu}U_\mu (n) ~- ~
\delta_{n,m+\mu}U_\mu^* (m)),\cr
\wilson\pm (n,m) & = {1\over 2} \sum_\mu
(2\delta_{n,m}~ - ~\delta_{m,n+\mu}U_\mu (n)~ - ~
\delta_{n,m+\mu}U_\mu^* (m))~\pm~ m\delta_{n,m} .\cr}
\eqno{(4.6)}$$
The parameter $m$ is restricted only by $0 < m < 2$.
The phases of the states $|L\pm>^{\rm WB}_U$ are chosen according to the
Wigner-Brillouin convention, i.e. ${}_1\!\! <L\pm|L\pm>^{\rm WB}_U$
is real and positive.
The determinant for a right handed Weyl fermion with $q_R=1$ is
$${}^{\rm WB}_{\ U}\!\!<R-|R+>^{\rm WB}_U
=\Bigl[{}^{\rm WB}_{\ U}\!\!<L-|L+>^{\rm WB}_U\Bigr]^* .\eqno{(4.7)}$$
The $|R\pm>^{\rm WB}_U$ are the highest energy eigenstates of  ${\cal H}^\pm$.
The regulated partition function for the 11112 model on the lattice 
becomes 
$$Z=\int [dU_\mu ] e^{S^w_g} 
\Bigl[{}^{\rm WB}_U\!\!<L-|L+>^{\rm WB}_U\Bigr]^4
{}^{\rm WB}_{\ U^2}\!\!<R-|R+>^{\rm WB}_{U^2} \eqno{(4.8)}$$

\bigskip
\centerline{\bf 5. Overlap when $\phi=0$}
\medskip

In this section we will show that the results of section 3 are reproduced
in the overlap formalism. To this end we consider the lattice form of
the uniform and singular gauge fields of section 3.
The link variables corresponding to the uniform field are
$$U^u_\mu(n)=e^{{2\pi i\over L} h_\mu}\eqno{(5.1)}$$
while the ones replacing the singular field are
$$\eqalign{
U^s_1(n_1,n_2) & = \cases{ e^{2\pi i h_1}
& if $n_1=0$ \cr 1 & elsewhere \cr} \cr
U^s_2(n_1,n_2) & = \cases{ e^{2\pi i h_2}
& if $n_2=0$ \cr 1 & elsewhere \cr} \cr
}\eqno{(5.2)}$$
$U^s_1 (0, n_2 )$ and $U^s_2 (n_1 , 0)$ represent 
the transition functions in the continuum.
We will denote the lattice 
overlap corresponding to a left-handed fermion by
$D_{\rm lat}^u(h_1,h_2;L)$ and by $D_{\rm lat}^s(h_1,h_2;L)$ for the uniform
and singular field respectively. 
For integers $n_1$ and $n_2$ it follows from (5.1) that 
$$D_{\rm lat}^u(h_1,h_2;L)
= D_{\rm lat}^u(h_1+n_1L,h_2+n_2L;L)\eqno{(5.3)}$$
whereas from (5.2) one concludes that 
$$D_{\rm lat}^s(h_1,h_2;L)
= D_{\rm lat}^s(h_1+n_1,h_2+n_2;L). \eqno{(5.4)}$$

$D_{\rm lat}^u(h_1,h_2;\infty)$ has no periodicity and reproduces
the continuum result $D^u(h_1,h_2)$ in (3.1).
In figure 5.1 we show how 
the limit $L\rightarrow\infty$ is approached. We have picked 
definite values for the $h_\mu$'s, 
namely $h_1=0.23$ and $h_2=0.37$. We then
evaluated  $D_{\rm lat}^u(0.23,0.37+n_2 ;L)$ as a function of $n_2$ for
two values of $L$, $L=20$ and $L=60$.
As one can see,
except for $n_2=\pm L/2$, all points lie on the continuum line given by
(3.2). The overlap for the uniform field approaches
the continuum result for all values of $h_\mu$.

\figure{5.1}{\captionfiveone}{5.0}
\figure{5.2}{\captionfivetwo}{5.0}
\figure{5.3}{\captionfivethree}{5.0}

On the other hand $D_{\rm lat}^s(h_1,h_2;L)$ 
is periodic in $h_\mu$
with period $1$ for all $L$. This is in agreement
with (3.4). We illustrate that
$D_{\rm lat}^u(h_1,h_2;\infty)$ and $D_{\rm lat}^s(h_1,h_2;\infty)$ reproduce
their continuum counterparts by an example. 
We fix $h_1=0.37$
and evaluate for various $L$'s 
the lattice overlaps as a function of $h_2$ in the range
$[0,0.5]$. The overlap for $-h_2$ is related to $h_2$ by
parity. In figure 5.2, we plot the locus of the complex
valued function $D_{\rm lat}^u(0.37,h_2;L)$ as $h_2$ is varied from $0$ to
$0.5$. We display $D^u(0.37,h_2 , L))$ 
at 11 equally spaced values of $h_2$, for
two values of $L$, $L=10$ and $L=18$.
The continuum function
$D^u(0.37,h_2)$ is also plotted. The lattice results
approach the correct continuum limit. 

Figure 5.3 is similar to figure 5.2, 
the singular $D_{\rm lat}^s(h_1,h_2;L)$ replacing the uniform
$D_{\rm lat}^u(h_1,h_2;L)$ and displaying 
one more value of $L$, $L=14$.
Again, the known continuum limit is
approached as $L$ increases. The lattice effects are larger for the singular field
than the uniform one. In particular the deviation from the
continuum has to be large when $h_2$ is close to
$0.5$ because the continuum answer has a discontinuity there.
On any finite lattice the discontinuity is approximated by a smooth function.
As the ultraviolet cutoff increases this smooth function has an
increasingly sharper profile around $h_2 =0.5$.

\bigskip

\centerline{\bf 6. Overlap along gauge orbits}
\medskip

The overlap formula for a single chiral fermion reproduces the
ordinary anomaly [\anom] for continuum gauge fields and therefore
is not gauge invariant. On the lattice there will be more terms,
of higher dimension, that also break the gauge symmetry. For
non-singular vector potentials the extra terms are local
and vanish as the lattice spacing goes to zero at a fixed
continuum external gauge field.  When we combine various 
fermions like in the 11112 model to make up an anomaly free theory,
the only
gauge breaking left on the lattice comes from these extra
terms.
Such extra gauge breaking terms will inevitably appear
in any approach where the fermion path integral
factorizes  into a product of factors, one for each
irreducible multiplet. This factorization
is a formal property of the continuum fermion path integral.
In our previous work [\nna]
we suggested dealing with the extra gauge breaking terms
by simply averaging over each gauge orbit. If they
are not too large, and
if anomaly free chiral gauge theories in the continuum
exist also beyond perturbation theory,
the most plausible outcome is that the averaging along the orbit
simply adds some irrelevant local gauge invariant terms to the
rest of the action. For example, a gauge breaking
term in an action for a pure gauge theory that has the form of
a mass term for the gauge bosons, when averaged
over the gauge orbits, induces only effects irrelevant in the
infrared, as long as its coefficient is not too large [\fnn].

In this section we
shall repeatedly start from some configuration that has a typical
gauge invariant content and average over its gauge orbit
by computing the overlap for many gauge transformations of
the original configuration. The overlap enjoys the nice property
that all the gauge breaking is restricted to
its phase [\nna]. Therefore, the gauge invariant
modulus of the overlap can be pulled out from the
integration along the gauge orbit and all we have to do
is to average the phases. The result of this averaging will be
some complex number, ${\cal Z}$.
$|{\cal Z}|$ might be
of some interest academically, but it can carry
only irrelevant information for the continuum target theory because
the modulus of the overlap already contains all the needed real part
of the effective action. We conclude that in practice one should
discard $\log |{\cal Z} |$, the contribution
of gauge averaging to the real part of the total fermion
induced effective action.
We only care about $\Phi$, where ${\cal Z} =
|{\cal Z}|e^{i\Phi}$.
$\Phi$ depends only on the gauge invariant content
of the initial configuration. Again a special property of the overlap
comes in here: Just like in the continuum, $\Phi$ will be parity
odd.
If after subtracting from $\Phi$ the
continuum value determined by the background we obtain a remainder
that admits an expansion in local operators, we have an
exact symmetry on the lattice that restricts these operators
to being parity odd. (Of course, such a local expansion is
an open possibility only if the perturbative anomalies cancel;
otherwise, gauge averaging the exponent of the
anomaly is bound to induce non-local terms.) There exist no
operators of dimension two or less that are parity
odd, local, and gauge invariant functionals of the background.
Assume that the continuum theory
is completely well defined and that
we start the gauge averaging from a relatively smooth background.
The dimensionalities of the allowed
operators imply that $\Phi$ will converge to the continuum gauge
invariant parity odd answer as we refine the lattice.

The simplest background one could imagine is certainly one in which
there is no curvature and all Polyakov loops are trivial.
The continuum phase $\Phi$ is supposed to vanish in this
background.
The lattice overlap on this ``trivial orbit'' ($\phi=0$ and $h_\mu=0$)
for a single chiral fermion was proven to be real
in [\nna].\footnote{*}{
In particular this nullifies the concerns expressed by
Golterman and Shamir in their publications dealing with the overlap.
Since the literature has become somewhat entangled, we set
things straight in this footnote to 
avoid confusing
those readers who are aware of most recent publications in the field,
but are not directly active in it.
The first paper by Golterman and Shamir, GSI, 
[Phys. Lett. B353 (1995) 84] criticizing the overlap
has two major errors. 
Both errors were anticipated in [\nna].
Since GSI incorrectly
claimed a rigorous mathematical equivalence between the overlap and a
specific waveguide model, we wrote a note [\nnb] 
repeating section 6 in [\nna]
which disproves this claim. We focused on only
one of the errors in GSI since it related to the central
result of GSI and showed that the rigorous
equivalence claimed in GSI was false.  After our note
[\nnb] was circulated, Golterman and Shamir published an erratum, 
GSII [Phys. Lett. B359 (1995) 422], 
in which they announced this error in GSI but claimed that
GSI still had shown that the overlap would fail. Specifically, their
claim was that, if one took a $U(1)$ chiral model in 2D with four
identical copies of its fermionic content and restricted the gauge
background to the trivial orbit, one would discover that the gauge
degrees of freedom labeling the points along the orbit will not
decouple. This statement was repeated in GSIII [hep-th/9509027]
and in other papers quoting GSI, GSII, GSIII.  
The statement is still
wrong and this goes back to the second major error 
in GSI and GSII which 
was anticipated in section 12 of [\nna]. 
There it was proven that
the overlap on the trivial orbit is real for any chiral fermion.
In particular, for the model considered in GSI, 
one has absolute independence of the points on the orbit.
This was pointed out in private to the authors before they
wrote their erratum GSII.}
Thus,
on the trivial orbit everything works fine
(there could be some sign switches, but they are very rare).
However, this might
cease being the case when we move away from the trivial orbit
because there the exact reality [\nna] no longer holds. We expect
the fluctuations of the overlap
along the orbit to gradually increase as the
background contains more local electric field
and less trivial Polyakov loop transporters.
We would like these
fluctuations to be insignificant in the continuum limit.
We shall work at a finite physical size in Euclidean
space-time. On the lattice the continuum limit is 
approached 
keeping $eL$ fixed in (4.3) and taking $L$ to infinity.
The physical size of the torus is proportional to $eL$. 
As discussed in the introduction, we focus on small systems. 

In this section
we evaluate the overlap for the 11112 model in various situations 
in order to study $\Phi$. We will show that if the $h_\mu$ are sufficiently small
the integration over gauge orbits yields the physical
result. If the $h_\mu$ are large, gauge violations of the type
discussed in sections 3 and 5 wash out the correct
value of $\Phi$ unless we exclude the ``singular'' gauge configurations.

Let us first consider the case when $h_\mu=0$. $eL$
controls the size of the fluctuations in $\phi$. 
If $eL$ is small, $\phi$
is close to zero and we expect the phase 
fluctuations along the orbit to
be small. These fluctuations are expected 
to also decrease as we take
$L$ to infinity, approaching continuum at fixed $eL$. 

\figure{6.1}{\captionsixone}{5.0}
\figure{6.2}{\captionsixtwo}{5.0}
\figure{6.3}{\captionsixthree}{5.0}

These trends are illustrated in
figures 6.1--6.3.
The overlaps for randomly drawn sets $\{ g(n) | n_1 , n_2 =1,...,L\}_i$
only differ by a phase.
Choosing the $g(n)$ randomly in $U(1)$, we computed this phase
and subtracted from
it the phase when all $g(n)=1$. When all $g(n)=1$
(Landau gauge) the phase is close to zero.
$\Phi$ vanishes in the continuum.
We denote the remainder phase for each set by $\Phi_i$.
The distribution of the $\Phi_i$ obtained
for $i=1,1000$ is then plotted.
The horizontal axis in all these plots is the value of the phase, $\Psi$,
in units of $\pi$. The vertical axis is the probability
of getting a particular
$\Psi$, $p(\Psi)$. We compute the phase for each chirality
individually. The left handed contribution is contained
in the points labeled by $1111$ and the right handed contribution
in the points labeled by $2$. The phase of
the product of the overlaps for the anomaly free combination is labeled
by $11112$. Only the $11112$ points are of direct relevance,
but the $1111$ and $2$ points indicate what an anomalous theory
would yield, and show how anomaly cancelation
works.
In figure 6.1 we have fixed $eL=0.1\pi$ and $L=6$.
In figure 6.2 we again have $eL=0.1\pi$ but $L=8$.
Each time, a $\phi$ was drawn using the distribution in (4.3)
which depends on $e=eL/L$.
The fluctuations in 6.2 have decreased somewhat relatively to 6.1. 
In figure 6.3 we have increased
$eL$ to $\pi$ and set $L=6$. Now the fluctuations are larger than in
figure 6.1.
However, the fluctuations in these three figures are all small and
all three distributions in each figure are well peaked around $0$.
Thus, figures 6.1--6.3 do not
exhibit any sizable numerical difference between anomalous and
anomaly free theories. This indicates that the specific lattice
gauge breaking effects, although not large in an absolute sense, are larger
numerically than the gauge breaking effects induced by the continuum anomaly,
fermion by fermion.

To attain an indication of what
kind of gauge transformations are responsible for
the larger effects we sample the points
on the orbit in a more controlled manner. We introduce
a parameterization $g(x)=e^{i\kappa\chi(x)}$
and draw the $\chi(x)$ from a distribution similar to (4.3):
$$S_{\rm gf} = {(eL)^2\over (e\pi)^2}
\sum_n [1- \cos ((\partial_1^*\partial_1
+ \partial_2^*\partial_2 )\chi(n))]\eqno{(6.1)}$$
This amounts to adding a $(\partial_\mu A_\mu)^2$ gauge fixing term
in the continuum.
Large momentum modes are suppressed
somewhat and $\kappa$ controls the amplitude.
As $\kappa$ is increased one spans the whole gauge orbit.
The added term to the action corresponds to a ``soft'' gauge
fixing term.
In principle we do not
want any ``soft'' or $\delta$--function gauge fixing at this stage
when the regularization method itself is tested.
However, when a simulation with the
objective of obtaining numerical results is carried out
keeping such gauge fixing terms might be useful. Here we added this term
just because it facilitates a better understanding of the difference between
anomalous and anomaly free theories. In figures 6.4 and 6.5 we plot
the phase distribution for two values of $\kappa$
at $eL=\pi$ and $L=6$. 
In figure
6.4, $\kappa=0.1$ and in figure 6.5, $\kappa=0.5$. The distribution for
the 11112 model is much sharper than the ones for 1111 or 2
illustrating the anomaly cancelation. In conclusion, the fluctuations
induced by the high momentum modes of the gauge transformations
are larger than the ones induced by the their low momentum modes.

\figure{6.4}{\captionsixfour}{5.0}
\figure{6.5}{\captionsixfive}{5.0}

Now we proceed to study the phase distribution when $h_\mu\ne 0$ to tie
in with the lack of gauge invariance observed in sections 3 and 5. If both
the $h_\mu$'s are sufficiently
small we do not expect large fluctuations. This is
illustrated in figures 6.6 and 6.7.
In both these figures we have set $\phi=0$ and the points in the orbit
are drawn randomly.
In figure 6.6 we have set $h_1=0.07$ and $h_2=0.13$ whereas in
figure 6.7 they are set at $h_1=0.23$ and $h_2=0.37$.
The dramatic difference between the two cases is obvious. No
qualitative change takes place 
if one also introduces a $\phi$ 
and this is shown in figure 6.8 where we
have generated a non-zero $\phi$ by setting $eL=\pi$ and kept the 
$h_\mu$'s at the same values as in figure 6.7.

\figure{6.6}{\captionsixsix}{5.0}
\figure{6.7}{\captionsixseven}{5.0}
\figure{6.8}{\captionsixeight}{5.0}

Figures 6.7 and 6.8 show that the distribution of phase is quite
biased and uneven if
the phase factors are large.
In section 3 and 5 we gave a specific example
whereby gauge invariance is violated in this case. Now we would like
to address the issue as to what class of gauge 
transformations are responsible
for the characteristics of the
observed distribution. In particular we would like to know if the
bias is all due to ``singular'' configurations of the type discussed
in sections 3 and 5.

\figure{6.9}{\captionsixnine}{5.0}
\figure{6.10}{\captionsixten}{5.0}

A first attempt is to introduce again a suppression of the high
momentum components of the gauge transformations as we did with
equation (6.1). This indeed removes the bias and unevenness, but is not
a detailed enough test because we want to see that
not all the randomness in the gauge degrees of freedom has
to be reduced in order to eliminate the gauge violations.
Randomness is needed for decoupling via the
F{\" o}rster--Nielsen--Ninomiya  [\fnn]
mechanism.
Only the very ordered large scale singularities are causing the
gauge
violations, and they are a feature of the
continuum theory, not the lattice.

To achieve this sharper distinction
we need a slightly different view of the basic idea
of lattice regularization. The traditional point of view is that
the $U_\mu$ variables correspond to path ordered integrals
of some vector potential defined globally over the lattice.
There exists another interpretation, due to Phillips [\patch].
The elementary cells of the dual lattice are viewed as a set
of patches
and the $U_\mu$ variables only represent the transition functions.
The connections inside the patches are thought of as being all trivial.
Which picture one adopts has no practical consequence in any
of the present day applications of lattice gauge theory.
In the context of chiral gauge theories however, the interpretation
matters conceptually. It makes sense, in particular in view of our
earlier discussion of the singular gauges, that a ``patchy''
view be maintained to some degree. It is not necessary to have
one distinct patch per elementary cell, nor do the link
variables have to all correspond to transition functions, or
all correspond to connections. It does make sense that the
sizes of the patches go to zero in the continuum so that
arbitrary gauge singularities are included. Thus we could imagine that
we have a certain number of lattice cells per patch
(this number could even be taken to diverge in the continuum,
but not as fast as the physical scales). Inside each patch
one could impose a smoothness condition on the gauge transformations,
or, even fix the gauge completely. This would be an
entirely local operation, different in essence from the usual
overall gauge fixing. The gauge transformations in distinct
patches would still be totally unrestricted.
In short, if the number
of lattice sites is ${\cal N}$ and the gauge group is ${\cal G}$,
the invariance group will no longer be ${\cal G}^{\cal N}$, but
${\cal G}^{{\cal N}^\prime}$ with ${\cal N}/{\cal N}^\prime$ being
some finite integer (or even diverging in the continuum).
Such
a setup would allow one to reduce the magnitude of the extra
lattice gauge breaking terms with no need to construct
continuum interpolations. Since the real part of the
lattice version of the chiral determinant can be made exactly
gauge invariant it is only the gauge transformations that might
need some control.

The above general point of view can be easily realized in our application.
We take an $L\times L$ lattice and
divide it into non-intersecting $L_b\times L_b$ blocks.
We think of each block
as
representing
a two dimensional contractible patch.
The links connecting the blocks 
represent transition
functions and the links inside the block represent a
gauge field in the local coordinates of that patch.
Now we set $\phi$ to zero
and consider two gauge realizations of the $h_\mu$'s (like in
section 5) as initial configurations.
As before, in (5.1-2), we refer to the first one as the ``uniform''
configuration and
the second one as the ``singular'' configuration.
Starting from the initial configuration we make gauge
transformations with $g(n)$
identical on every site in a single block
but taking random and independent
values on distinct blocks.
We picked $h_1=0.23$ and $h_2=0.37$. The chiral determinant for the 11112
model in this background is complex when we are in the ``uniform'' gauge,
i.e. $\Phi \ne 0$.
We computed the distribution of the difference phases $\Phi_i$
when drawing random gauge transformations constrained per block
as described above. In figure 6.9 we plot the results when
the starting point was
a ``uniform'' configuration and in figure 6.10 we plot
the distribution when the starting point was a ``singular''
configuration.
In both figures $L=9$ and $L_b=3$.
Clearly the fluctuations in the ``singular'' case are qualitatively
different. 
The gauge transformations we included did allow many singularities 
``between'' the blocks. However, the restrictions did not allow
the gauge transformations to ``pile up'' all of the Polyakov loop
transporters
on a single link (or very few links) in the uniform case. As
a result figure (6.9) shows no significant gauge violations.
However, in the singular case the restricted gauge transformations
are allowed to ``spread out'' the Polyakov loops over many
different block boundaries,
(the location of the singularities in the initial
configuration was picked at an ordered
set of block boundaries compatible with (5.2)).
Thus figure (6.10) does exhibit 
a significant amount of gauge violation.
We conclude that there is evidence on the lattice
that the real source for all important gauge violations
is the one identified in the continuum and that there is
no reason to suspect the lattice overlap of introducing new
undesirable effects.

We now wish to show that gauge averaging along the orbits would
perform satisfactorily for any background if we could forbid
the singular gauge transformations.
Since the electric field never caused any difficulties
we consider only the
case of large $h_\mu$'s but with $\phi=0$. We
restrict ourselves to a uniform initial configuration
and do blocked gauge transformations as above. The phase
fluctuations are controlled but larger than in the other cases of
interest, namely $h_\mu=0$ or small $h_\mu$  with non-zero $\phi$.
We fixed $L_b=3$ and considered $L=6,9,12,15,18,21$. In each
of these lattices we considered $10,000$ points
on the orbit of blocked gauge transformations
and averaged as before. The
results for ${\rm Im} (\log <e^{i\pi \Phi_i }>) /\pi$
are tabulated in Table 6.1. The result for the 1111 or the 2 case
appear to grow as $L$ is increased whereas the result for the 11112 case
is stable at some small value. Increasing the block size further decreases
this small value.
We see that indeed anomaly cancelation does play a role, as the
right and left sectors by themselves do not seem to allow the lattice
$\Phi$ to approach any limit, and definitely not the correct
one.
This indicates a restoration of gauge
invariance in the 11112 model as we go to the continuum limit.
The gauge transformations we included are very
random on scales larger than the block size
but they cannot bring the Polyakov loops to the ``singular'' gauge.
While this proves our point we would not go as far as suggesting
this as a possible redefinition of the 11112 model, simply because
the decision about which initial configuration to pick
involves a nonlocal choice.
\bigskip
\centerline{Table 6.1}
\medskip
{\null\hfill\vbox{\offinterlineskip
\hrule
\halign{&\vrule#\hfil&
  \strut\quad#\hfil\quad\cr
height2pt&\omit&&\omit&&\omit&&\omit&\cr
&$L$&&$1111$&&$2$&&$11112$&\cr
height2pt&\omit&&\omit&&\omit&&\omit&\cr
\noalign{\hrule}
height2pt&\omit&&\omit&&\omit&&\omit&\cr
&$6$&&$0.032\pm 0.007$&&$-0.134\pm 0.008$&&$-0.109\pm 0.014$&\cr
&$9$&&$0.147\pm 0.007$&&$-0.207\pm 0.008$&&$-0.089\pm 0.019$&\cr
&$12$&&$0.171\pm 0.013$&&$-0.250\pm 0.004$&&$-0.093\pm 0.019$&\cr
&$15$&&$0.215\pm 0.014$&&$-0.289\pm 0.011$&&$-0.083\pm 0.031$&\cr
&$18$&&$0.213\pm 0.023$&&$-0.310\pm 0.012$&&$-0.098\pm 0.044$&\cr
&$21$&&$0.238\pm 0.016$&&$-0.308\pm 0.009$&&$-0.083\pm 0.042$&\cr
height2pt&\omit&&\omit&&\omit&&\omit&\cr
}
\hrule}\hfill}
\bigskip

\centerline {\bf 7. A new 11112 model}
\medskip
Up to now we restricted our attention to the 
11112 model on the torus
with antiperiodic boundary conditions for all fermions.
This model has an $SU(4)$ flavor symmetry. 
We identified gauge orbits on 
which gauge invariance under certain singular gauge 
transformations cannot be maintained.

In this section we show that by allowing the fermionic 
boundary conditions to 
break the flavor $SU(4)$ we can arrange for 
both $Z^u (h_1 , h_2 )$ and $Z^s (h_1 , h_2 )$ 
(see (3.5) and (3.7)) to be 
periodic under $h_\mu \rightarrow h_\mu +{1\over 2} n_\mu$
with integer $n_\mu$. One can then restrict the $h_\mu$ to the
interval $[ -{1\over 4}, {1\over 4}]$ ensuring
$Z^u (h_1 , h_2 ) = Z^s (h_1 , h_2 )$ for all $h_\mu$ (see (3.9)).

The new 11112 model 
easily generalizes to models containing one 
left handed fermion 
of charge $Q$ and $Q^2$ right handed fermions of unit charge. 
We define the
new models for arbitrary integer $Q > 0$.
The left handed fermion, $\psi_L$ obeys anti-periodic boundary conditions
as before. It is convenient to use a doubled index to label the 
different flavors of the righthanded fermions:
$\psi_{R, \alpha\beta}$ where $\alpha,\beta =1,2,...,Q$. The boundary
conditions for the $\psi_R$ are
$$\eqalign{
\psi_{R,\alpha\beta} (x_1 + l N_1 , x_2 +l N_2 ) &=
e^{{{2\pi i} \over Q}(N_1\alpha +N_2 \beta -{{N_1 +N_2 }\over 2}) } 
\psi_{R,\alpha\beta} (x_1 , x_2 ) \cr
&=(-1)^{N_1 + N_2 } e^{{{2\pi i}\over Q} [N_1 (\alpha - {{1+Q}\over 2})+
N_2 (\beta -{{1+Q}\over 2})]} \psi_{R,\alpha\beta} (x_1 , x_2 ) \cr}
\eqno{(7.1)}$$

In the old model the generalization of (3.5) to the charge $Q$
case would have been:
$$
Z^u ( h_1 , h_2 ) =[D^u (h_1 , h_2 )]^Q [D^u (Qh_1 , Qh_2 )]^* 
\eqno{(7.2)}$$
In the new model the above is replaced by
$$
Z^u_{\rm new} (h_1 , h_2 ) =\prod_{\alpha , \beta =1}^Q
\left [ D^u ( h_1 + {\alpha \over Q} - {{1+Q}\over {2Q}} \ , \
h_2 + {\beta \over Q} - {{1+Q}\over {2Q}} ) \right ] ~
\left [D^u (Qh_1 , Qh_2 ) \right ]^*
\eqno{(7.3)}$$

Using the product representations of (3.1) it is not 
hard to prove that
$$
\prod_{\alpha ,\beta =1}^Q
\left [ D^u ( h_1 + {\alpha \over Q} - {{1+Q}\over {2Q}} \ , \
h_2 + {\beta \over Q} - {{1+Q}\over {2Q}} )
\right ] = D^u (Qh_1 , Qh_2 )
\eqno{(7.4)}$$
Hence
$$
Z^u_{\rm new} (h_1 , h_2 )=|D^u (Qh_1 , Qh_2 )|^2\eqno{(7.5)}$$

It is now obvious that $Z^u_{\rm new} (h_1 , h_2 )=
Z^s_{\rm new} (h_1 , h_2 )$ since one has periodicity under
independent shifts of each $h_\mu$ by integral multiples of $1/Q$.
Moreover, all the imaginary parts of the continuum action vanish just
like in a vectorial model. Nevertheless, the model is chiral
as evidenced by the different charges carried by the left and right
handed fermions. The model is definitely
not vectorial even ``in disguise''
when $Q$ is even since then the total number of Weyl
fermions is odd.

On the lattice, the  relevant product of overlaps will contain
an imaginary part since the continuum cancelation between the right
and left handed fermions no longer holds exactly. The unwanted,
gauge breaking terms discussed in the previous section are present
here also since the theory is not vectorial. If the overlap
indeed produces correct results in the continuum limit after gauge
averaging along the orbits we should see no longer any significant
difference between backgrounds with relatively small values of $h_\mu$
and backgrounds with larger values. In particular, the
qualitative difference between figures 6.6 and 6.7 in the
previous section should disappear.

\figure{7.1}{\captionsevenone}{5.0}
\figure{7.2}{\captionseventwo}{5.0}

To check this we set $Q=2$ as before. We then carried out
numerical simulations similar to the ones that produced figures 6.6 and 6.7 but
with the new model. The results are in figures 7.1 and 7.2,
where 7.1 is the analogue of 6.6 and 7.2 replaces 6.7.
While 6.6 and 7.1 are similar, 6.7 and 7.2 are quite different.
In 7.2 we see clear evidence that the phases coming from the
left handed and from the right handed fermion cancel during gauge averaging 
leaving us with an answer that is real (up to small
ultraviolet and statistical effects).\footnote{*}{
We also looked at the contribution to the real part of the effective action -- which we discard -- 
and note that in the new model the
induced real part is numerically smaller then in the old model.}

Therefore, the new chiral 11112 model is gauge invariant in the
continuum under all gauge transformations on all gauge orbits and 
gauge averaging of the
WB phases of the lattice overlap 
reproduces the expected results.
We checked that turning on electrical 
fields again has no
significant effect on the behavior of gauge averaging. 
Armed with this
evidence we see nothing qualitative left to check regarding the applicability of the overlap
formalism to the zero 
topology sector of this new 11112 model.
Recalling our study of the Schwinger model [\nnv] we consider it very likely
that no difficulties will be uncovered at non-zero topology.

We conclude that the new 11112 model very
likely has the following two
main properties:
\item{$\bullet$} It is an exactly
soluble chiral model with $U(1)$ gauge
interactions and fermion number violation.
\item{$\bullet$} The exact continuum results 
are reproduced by the overlap formalism.

To prove the above one would have to carry out a
complete numerical simulation. This simulation would be similar
to the ones in [\nnv]. 
The expectation values of the
appropriate 't Hooft vertex ( $\prod_{\alpha,\beta =1}^Q
[\bar\psi_{R,\alpha\beta} (x) ] [ ( \psi_L \partial_L \psi_L ) (x)]$ )
obtained on a sequence of
increasingly finer lattices representing a fixed physical size
would have to be extrapolated to the continuum limit.
The simulation would employ gauge averaging,
but this would have almost
no numerical effect since the real part of
the fermion induced gauge action would
be discarded.
In order to know whether the result is correct, one would
need the exact continuum expressions for the 't Hooft
vertices in the appropriate finite Euclidean volume. The needed formulae
would have to be obtained following [\sw].

\bigskip
\centerline{\bf 8. Conclusions}
\medskip

Surprisingly,
chiral abelian two dimensional gauge theories on a torus do not provide
a completely
transparent testing ground for the overlap lattice formalism. More specifically,
the ingredient of this formalism one would like to test 
outside perturbation theory is the
Wigner--Brillouin phase choice. All other 
non-perturbative ingredients can and have been
tested in vectorial theories with global chiral symmetries. 
These tests
were all successful. In the present work we
focused on two versions of
the 11112 chiral $U(1)$ model. What we found
supports the validity of the basic strategy of [\nna].
This strategy employs the Wigner--Brillouin phase choice and
restores exact gauge invariance by averaging along gauge
orbits. It is logically based on the work of
F{\" o}rster--Nielsen--Ninomiya [\fnn].

Let us first list our findings without our interpretation.

\leftline{\sl A) 11112 with antiperiodic boundary conditions for
all fermions.}

\item{$\bullet$} There exists a large class of gauge orbits,
not of zero measure, on which the gauge averaging of W-B phases
appears to wash out the expected gauge invariant, parity odd,
continuum answer 
one would obtain in a smooth background.
\item{$\bullet$} This class can be identified in the continuum,
independently 
of any knowledge about the intended regularization
procedure.
\item{$\bullet$} On each of the above ``bad'' orbits the source
of the wash-out was identified. We found that the ``bad''
gauge transformations on the ``bad'' orbits approximate
continuum $A_\mu$'s that have gauge singularities in the form
of coherently ordered linear delta functions.
\item{$\bullet$} There also exists a large class of ``good'' gauge
orbits, again not of zero measure, on which the gauge averaging
of W-B phases produces the expected gauge invariant, parity odd,
continuum answer. The gauge averaging on the ``good'' orbits
is unrestricted, including gauge transformations that would be
deemed ``bad'' on ``bad'' orbits.
\item{$\bullet$} On ``bad'' orbits, if one eliminates the ``bad''
gauge transformations, gauge averaging does produce the expected answer.
Anomaly cancelation is essential. 
\item{$\bullet$} If one adds a sufficiently strong soft gauge breaking
term of the type $\int (\partial_\mu A_\mu )^2$ and then uses gauge
averaging along orbits all indications are that the continuum
results will be fully reproduced by the overlap.

\leftline{\sl B) 11112 with special fermionic boundary conditions.}

\item{$\bullet $} All difficulties in the continuum
and on the lattice disappear. Gauge averaging works fine.

Our interpretation, based on a fiber bundle view of the domain
of integration in the path integral, is that the ``bad'' gauge
transformations spoiling gauge invariance on the ``bad'' orbits
are a continuum difficulty. This view might be contested
if one insists that all the allowed vector potentials
in the zero topology sector are essentially smooth functions in the continuum.

Our message for the future is two-fold: The first part is independent
of our interpretation. One should carry out a
complete dynamical simulation of model {\sl B)} and check whether the
continuum results are reproduced quantitatively.
The new fermion boundary conditions remind us 
of the twisted many flavor Schwinger models 
devised by Shifman and Smilga [\shsm]. 

The second part of our message is that attention should be paid
to gauge singularities that ordinary fiber bundle representations
admit. A thorough investigation of chiral fermions in such backgrounds
should be possible by continuum methods and might reveal differences between
chiral and non--chiral anomaly free theories.

One may say that the evidence in favor of the overlap
formalism accumulated to date is enough to turn
the tables around and accept this formalism as a basis for the investigation
of the existence of asymptotically free chiral gauge theories
outside perturbation theory.
If the gauge breaking terms turn out to be too strong
in some particular case one can
always use patches like at the end of section 6 to weaken them.
If this does not work one should suspect not only
the regularization but entertain the possibility
that the fault is in the continuum.

\noindent {\bf Acknowledgments}:
R. N. was supported in part by the DOE under grant \# DE-FG06-91ER40614
and under grant \# DE-FG06-90ER40561.
H. N. was supported in part by the DOE under grant \#
DE-FG05-90ER40559.

\vskip .1in
\noindent{\bf References}
\vskip .2in
\item{[\inter]} M. G\"ockler, A.S. Kronfeld, G. Schierholz and U.J. Wiese,
Nucl. Phys. B404 (1993) 839; M. G\"ockler, private communication.
\item{[\sw]} I. Sachs and A. Wipf, Helv. Phys. Acta 65 (1992) 652.
\item{[\nna]}  R. Narayanan, H. Neuberger, Nucl. Phys. B443 (1995) 305.
\item{[\fnn]} D. Foerster, H.B. Nielsen and M. Ninomiya,
Phys. lett. B94 (1980) 135.
\item{[\detu]} L. Alvarez-Gaume, G. Moore and C. Vafa, Comm. Math. Phys.
6 (1986) 1; R. Narayanan and H. Neuberger, Phys. Lett. B348 (1995) 549;
C.D. Fosco and S. Randjbar-Daemi, Phys. Lett. B354 (1995) 383; 
C.D. Fosco, hep-th/9511221. 
\item{[\anom]}  R. Narayanan, H. Neuberger,
Phys. Rev. Lett. 71 (1993) 3251; Nucl. Phys. B412 (1994) 574;
S. Randjbar-Daemi and J. Strathdee, Phys. Rev D51 (1995) 6617;
Nucl. Phys. B443 (1995) 386; Phys. Lett. B348 (1995) 543;
hep-th/9512112.
\item{[\nnb]} R. Narayanan and H. Neuberger, Phys. Lett. B358 (1995) 303.
\item{[\patch]} A. Phillips, Ann. of Phys. 161 (1985) 399.
\item{[\nnv]} R. Narayanan, H. Neuberger and P. Vranas, 
Phys. Lett. B353 (1995) 507; hep-lat/9509047.
\item{[\shsm]} M.A. Shifman and A.V. Smilga, Phys. Rev. D50 (1994) 7659.
\vfill
\eject
\end